\documentclass[conference]{IEEEtran}
\IEEEoverridecommandlockouts
\usepackage{cite}
\usepackage{algorithm2e}
\usepackage{algorithmic}
\usepackage{graphicx}
\usepackage{array, makecell} 
\usepackage{diagbox}
\usepackage{subcaption,multicol}
\usepackage{multirow}

\usepackage{epstopdf}
\usepackage{epsfig}
\usepackage{float}
\usepackage{enumitem}
\newcommand{\alg}{\texttt{CSE-FSL} }
\newcommand{\algg}{\texttt{CSE-FSL}}

\usepackage{amssymb,amsmath,amsfonts,bm,epsfig,graphicx,theorem}
\usepackage[dvipsnames]{xcolor}
\newcommand{\comment}[1]{}

\usepackage{comment}

\newcommand{\av}{\mathbf{a}}

\newcommand{\mv}{\mathbf{m}}

\newcommand{\xv}{\mathbf{x}}

\newcommand{\yv}{\mathbf{y}}
\newcommand{\zv}{\mathbf{z}}

\newcommand{\argmin}{\mathop{\mbox{\rm arg\,min}}}

\newcommand{\vect}[1]{\mathbf{#1}}

\newcommand{\norm}[1]{\left \| #1 \right \|}

\newcommand{\Expect}{\mathbb{E}}

\newcommand{\qth}[1]{\left[ #1 \right]}
\newcommand{\sth}[1]{\left\{ #1 \right\}}

\newtheorem{Proposition}{Proposition}

\newtheorem{Assumption}{Assumption}

\ifodd 1
\newcommand{\congr}[1]{{\color{red}#1}}
\else
\newcommand{\congr}[1]{#}
\fi

\ifodd 1
\newcommand{\congc}[1]{{\color{red}(Cong: #1)}}
\else
\newcommand{\congc}[1]{}
\fi

\newcommand{\mypara}[1]{{\smallskip \noindent \bf #1}\hspace{0.1in}}

\def\BibTeX{{\rm B\kern-.05em{\sc i\kern-.025em b}\kern-.08em
    T\kern-.1667em\lower.7ex\hbox{E}\kern-.125emX}}

\begin{document}

\title{Communication and Storage Efficient\\
Federated Split Learning
\thanks{The work is partially supported by the National Science Foundation under Grant CNS-2002902, and the Commonwealth Cyber Initiative (CCI) of Virginia under Award VV-1Q23-005.}

}

\author{\IEEEauthorblockN{Yujia~Mu and Cong~Shen}
\IEEEauthorblockA{Charles L. Brown Department of Electrical and Computer Engineering\\ University of Virginia\\
Charlottesville, VA, USA}
}

\maketitle

\begin{abstract}

Federated learning (FL) is a popular distributed machine learning (ML) paradigm, but is often limited by significant communication costs and edge device computation capabilities. Federated Split Learning (FSL) preserves the parallel model training principle of FL, with a reduced device computation requirement thanks to splitting the ML model between the server and clients. However, FSL still incurs very high communication overhead due to transmitting the smashed data and gradients between the clients and the server in each global round. Furthermore, the server has to maintain separate models for every client, resulting in a significant computation and storage requirement that grows linearly with the number of clients. This paper aims at solving these two issues by proposing a communication and storage efficient federated split learning (\texttt{CSE-FSL}) strategy, which utilizes an auxiliary network to locally update the client models while keeping only a single model at the server, hence avoiding the communication of gradients from the server and greatly reducing the server resource requirement. Communication cost is further reduced by only sending the smashed data in selected epochs from the clients. We provide a rigorous theoretical analysis of \alg that guarantees its convergence for non-convex loss functions. Extensive experimental results demonstrate that \alg has a significant communication reduction over existing FSL techniques, while achieving state-of-the-art convergence and model accuracy, using several real-world FL tasks.

\end{abstract}


\section{Introduction}
\label{sec:intro}
As an emerging distributed machine learning (ML) paradigm, federated learning (FL) \cite{mcmahan2017communication} allows clients to collaboratively train ML models without uploading their sensitive data to the server. While the FL framework helps alleviate the data privacy concern, most existing FL algorithms rely on that the clients have \emph{sufficient computation and storage resources} to perform local updates on the ML models, especially deep neural networks (DNNs). However, when the computing power and memory of the clients are limited (e.g. edge devices), FL is infeasible to handle large models.

Split learning (SL) \cite{gupta2018distributed} proposes to split the DNN model between client and server to address this issue. The first few layers are trained at the client, while the remaining are only stored and updated at the server. Since each client only needs to store and train the first few layers of the model, the storage and computational burden on clients is reduced. Nevertheless, one major limitation of SL is the significant time delay since a ML model is trained across multiple clients sequentially. 

Federated split learning (FSL) \cite{thapa2020splitfed} combines the strengths of FL, which is parallel processing among distributed clients, and the advantages of SL, which is model splitting between clients and server during training. Accordingly, FSL has less training time than SL and reduces the storage and processing load for resource-limited devices over FL. Nevertheless, the underlying model partitioning leads to increased communication cost. Specifically, the communication burden can be substantial for transmitting the forward signals (smashed data) and backward signals (gradients) in each global round. One solution to reducing the communication cost is the local loss-based training \cite{han2021accelerating}, by updating the client-side model locally without waiting for receiving the gradients from the server. However, this architecture is most suitable for scenarios in which the server has enough storage and computing power, because the resource consumption of the server is proportional to the number of clients. Correspondingly, FSL does not scale well with the number of clients. 

The goal of this work is to make FSL communication and storage efficient, so that its practicality can be improved to a level that propels its adoption in massive resource-constrained devices while managing the storage and computation requirement at the server. Towards that end, we propose a novel communication and storage efficient federated split learning technique (\texttt{CSE-FSL}), which not only greatly reduces the massive communication cost but also drastically saves storage by keeping a \emph{single} server-side model regardless of the number of clients. 
Given a mini-batch of data, the client does not need to communicate per-batch forward signals to the server thanks to the adoption of auxiliary networks. On the server side, we propose a single model training strategy that performs model updates only when the smashed data from many clients are received. The server then uses the data from different clients \emph{sequentially}, mimicking a multi-epoch training. Our method significantly improves the communication efficiency while reducing the amount of data communicated in both uplink and downlink. Before each aggregation, all clients send their locally trained client-side model and auxiliary network to the server. Then all the received models are aggregated at the server and redistributed to clients. These aggregated models are used as the initial model for the next round of \texttt{CSE-FSL}. 

We provide rigorous theoretical analysis to guarantee the convergence of \alg with non-convex loss functions, which helps understand the interplay among key system constraints and hyperparameters in the convergence process.  We validate these theoretical results by performing real-world FL tasks for both  independent and identically distributed (IID) and non-IID cases on two widely adopted datasets, CIFAR-10 \cite{krizhevsky2009learning} and F-EMNIST \cite{caldas2018leaf}. Experimental results show that \alg significantly outperforms existing FSL solutions with a single model or multiple copies on the server. 

The remainder of this paper is organized as follows. 
Existing FSL methods and their issues are described in Section~\ref{sec:model}. 
The proposed \alg method, as well as the convergence analysis, are presented in Section~\ref{sec:CSE-FSL} . Experimental results are given in Section~\ref{sec:sim}, followed by the conclusions in Section~\ref{sec:conc}.


\section{Existing Federated and Split Learning}
\color{black}
\label{sec:model}
We begin by presenting the underlying optimization problem, and then describe the standard federated and split learning pipeline. 
We then discuss the limitations of existing federated and split learning methods.
\color{black}

\subsection{Distributed SGD}
\label{sec:modelFL}

We study the standard empirical risk minimization (ERM) problem in ML:
\begin{equation} \label{eqn:erm}
\min_{\xv \in \mathbb{R}^d} F(\xv) = \min_{\xv \in \mathbb{R}^d} \frac{1}{|D|} \sum_{\zv \in D} l(\xv; \zv),
\end{equation}
where $\xv \in\mathbb{R}^d$ is the ML model variable that one would like to optimize, $l(\xv; \zv)$ is the loss function evaluated at model $\xv$ and data sample $\zv=(\vect{z}_{\text{in}}, z_{\text{out}})$ describing an input-output relationship of $\vect{z}_{\text{in}}$ and its label $z_{\text{out}}$, and $F: \mathbb{R}^d\rightarrow\mathbb{R}$ is the differentiable loss function averaged over the total dataset $D$. We assume that there is a latent distribution $\nu$ that controls the generation of the global dataset $D$, i.e., every data sample $z \in D$ is drawn independently and identically distributed (IID)\footnote{In  Section~\ref{sec:sim} we will numerically evaluate non-IID datasets.} from $\nu$.  We denote
$
    \xv^* \triangleq \argmin_{\xv \in \mathbb{R}^d} F(\xv), f^* \triangleq F(\xv^*).
$

One category of distributed and decentralized ML, including FL, aims at solving the ERM problem \eqref{eqn:erm} by using a set of clients that run local computations in parallel, hence achieving a wall-clock speedup compared with the centralized training paradigm. We consider a distributed ML system with one central parameter server (e.g., at the base station) and a set of $n$ clients (e.g., IoT devices).  Mathematically, problem \eqref{eqn:erm} can be equivalently written as
 \begin{equation} \label{eqn:erm2}
\min_{\xv \in \mathbb{R}^d} F(\xv) = \min_{\xv \in \mathbb{R}^d} \frac{1}{N} \sum_{i=0}^{N-1}  F_i(\xv),
\end{equation}
where $F_i(\xv)$ is the local loss function at client $i$, defined as the average loss over its local dataset $D_i$: $F_i(\xv) =   \frac{1}{|D_i|} \sum_{\zv \in D_i}  l(\xv; \zv)$,
We make the standard assumption that local datasets are disjoint, and $D = \cup_{i \in [n]} D_i$. This work largely focuses on the \emph{full clients participation} setting, where all $N$ clients participate in every round of distributed SGD. We will report numerical results for partial clients partipation in Section~\ref{sec:sim}. 
To ease the exposition and simplify the analysis, we also make the assumption that all clients have the same size of local datasets, i.e., $|D_i|=|D_j|, \forall i, j \in [n]$.

\subsection{Federated and Split Learning}
\label{sec:modelFSL}

\begin{figure}
    \centering
    \includegraphics[width=0.48\textwidth]{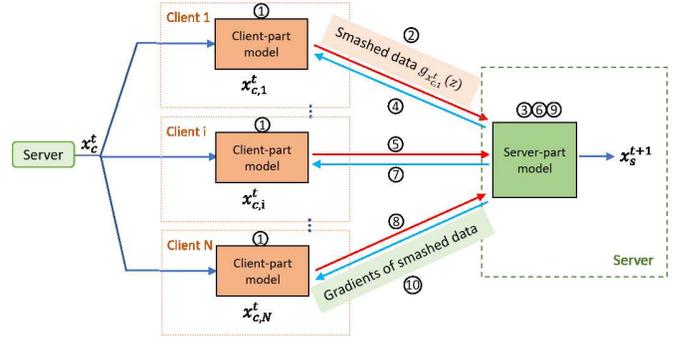}
    \caption{\small Traditional FSL pipeline in the $t$-th global round.}
    \label{fig:FSLDiagram}
\end{figure}

We describe the original SplitFed framework \cite{thapa2020splitfed}, which we closely follow, and explicitly explain how to train client-side models in parallel (the federated learning component). 
The overall diagram is depicted in Fig.~\ref{fig:FSLDiagram}. We first split the complete model $\xv$ into the client-side model $\xv_c$ and the server-side model $\xv_s$. Then, all clients download the initial client-side model from the server and carry out the forward propagations on their client-side models in parallel, before uploading their smashed data (activations) and the corresponding labels to the server. Next, the server continues to process the forward propagation and back-propagation on its server-side model sequentially with respect to the smashed data and then updates the server-side model. For each model update, the server finally sends the gradients of the smashed data to the respective client for their back-propagation and model update. These steps are repeated until all training data is processed once (one epoch) and the clients upload the updated client-side model to the server. Finally, the server aggregates them to generate a global client-side model and then redistributes to the clients for the next round. We note that this is the standard FSL workflow in the literature.

\subsection{Large communication costs}
\label{subsec:largecost}
From the FL perspective, FSL is a framework of training the client-side models in parallel, which improves the efficiency of local model training compared with SL. However, existing FSL methods all suffer from significant communication costs. For each mini-batch data samples, the server must collect smashed data from all participating clients to train the server-side model (upstream), and all participating clients need to wait for gradients from the server to update their local models (downstream). As a result, the communication overhead for both upstream and downstream require significant resources at each global round.

\color{black}

\section{Communication and Storage Efficient Federated Split Learning}
\color{black}
\label{sec:CSE-FSL}


We propose a novel communication and storage efficient federated split learning technique (\texttt{CSE-FSL}), which not only reduces both upstream and downstream communication costs but also saves storage by keeping a \emph{single} server-side model. Next, we will introduce the auxiliary network and the client-side and server-side loss functions, and then present details of \algg.
\color{black}

\subsection{Auxiliary Network}
\label{subsec:auxiliary}

In conventional FSL, the client-side model is updated with the backpropagated signals from the server. In fact, the signals are the gradients of the smashed data, which are obtained by calculating the loss from the server-side model. To break this loop, \cite{han2021accelerating} adds an auxiliary network $\av_c$ to the client-side model, and uses it to calculate the local loss. In other words, the output of the client-side model is the input of the auxiliary network. Both convolutional neural networks (CNN) and multi-layer perceptrons (MLP) can be utilized as the auxiliary network. The impact will be discussed in the journal version due to space limitation. 

For the clients, the goal is to find $\xv_c$ and $\av_c$ that solve the ERM problem \eqref{eqn:erm}:
\begin{equation} \label{eqn:erm_c}
\min_{\xv_c, \av_c} F_c(\xv_c) = \min_{\xv_c, \av_c} \frac{1}{N} \sum_{i=0}^{N-1}  F_{c,i}(\xv_c, \av_c),
\end{equation}
where $F_{c,i}(\xv_c, \av_c)$ is the local loss function at client $i$, defined as the average loss over its local dataset $D_i$: $F_{c,i}(\xv_c, \av_c) =   \frac{1}{|D_i|} \sum_{\zv \in D_i}  l(\xv_c, \av_c; \zv) $.

With the auxiliary network, the clients can update the models locally without waiting for the transmission of the gradients of the smashed data. For the server, however, \cite{han2021accelerating} requires significant storage space because it keeps one individual copy of server-side model for every client, which does not scale with the number of clients and size of the model. 

The goal is to find $\xv_s$ that solves \eqref{eqn:erm} based on the optimal client-side model $\xv_c^*$ defined in \eqref{eqn:erm_c}:
\begin{equation} \label{eqn:erm_s}
\min_{\xv_s} F_s(\xv_s) = \min_{\xv_s} \frac{1}{N} \sum_{i=0}^{N-1}  F_{s,i}(\xv_s, \xv_c^*),
\end{equation}
where $F_{s,i}(\xv_s, \xv_c^*)$ is the local loss function of current server-side model corresponding to the dataset of clients $i$, defined as the average loss over its local dataset $D_i$: $F_{s,i}(\xv_s, \xv_c^*) =   \frac{1}{|D_i|} \sum_{\zv \in D_i}  l(\xv_s ; g_{x_c^*}(\zv))$. Note that the smashed data of the optimal client-side model $x_c^*$ with input $\zv \in D_i$ is denoted by $g_{x_c^*}(\zv)$.

\subsection{\texttt{CSE-FSL}}
\label{subsec:CSE-FSL}
\begin{figure}
    \centering
    \includegraphics[width=0.5\textwidth]{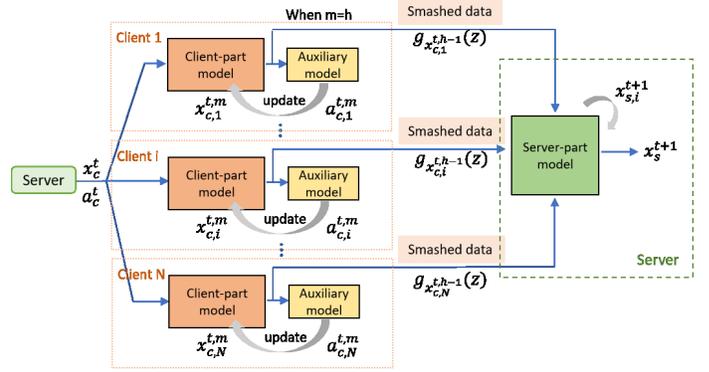}
    \caption{\small End-to-end \alg pipeline in the $t$-th global round.}
    \label{fig:SystemDiagram}
\end{figure}

\color{black}
In the proposed solution, we also consider an auxiliary network in the client-side model, but keep only a single server-side model instead of multiple models to reduce the storage and computing requirement from $O(N)$ to a constant. Additionally, to further reduce the communication cost, the clients in our method do not upload smashed data in each mini-batch training. Instead, the server updates the model in every $h$ batches of data, and denote the initial model as $\{\xv_c^0, \xv_s^0, \av_c^0\}$. The overall system diagram is depicted in Fig.~\ref{fig:SystemDiagram}.  In particular, the pipeline works by iteratively executing the following steps at the $t$-th learning round, $\forall t \in [T] \triangleq \sth{0, 1, \cdots, T-1}$.

\label{sec:algorithm}
\mypara{Step 1: model download.} At the beginning of global round $t$, client $i \in [N]$ downloads the client-side model $\xv_c^t$ and the auxiliary model $\av_c^t$ from server, and sets $\xv_{c,i}^{t,0} = \xv_c^t, \av_{c,i}^{t,0} = \av_c^t$.

\addtolength{\topmargin}{0.02in}

\mypara{Step 2: feedforward and smashed data upload.} For each mini-batch of training sample $\zv \in \tilde{D_i}$ (note that $\tilde{D_i} \subset D_i, D = \cup_{i \in [n]} D_i$), each client $i$ (in parallel) performs feedforward to the last layer of the auxiliary network $\av_{c,i}^{t,m}$ based on the client-side model $\xv_{c,i}^{t,m}$. In this process, we can calculate the local loss  $F_{c,i}(\xv_{c,i}^{t,m}, \av_{c,i}^{t,m})$ (see Step 3) for all training samples $\zv \in \tilde{D_i}$. Note that if current batch number $m$ satisfies $m \bmod h=0$, each client $i$ computes the smashed data $g_{\xv_{c,i}}(\zv)$, which is the output of the client-side model, and uploads the smashed data and the labels corresponding to the batch data to the server.

\mypara{Step 3: model update.} Based on the local loss from Step 2, the client-side model and the auxiliary network can be updated through backpropagation:

 \begin{equation} \label{eqn:step3}
\begin{cases}
      \xv_{c,i}^{t,m+1} = \xv_{c,i}^{t,m} - \eta_t \tilde{\nabla}_x F_{c,i}(\xv_{c,i}^{t,m}, \av_{c,i}^{t,m}) \\
      \av_{c,i}^{t,m+1} = \av_{c,i}^{t,m} - \eta_t \tilde{\nabla}_a F_{c,i}(\xv_{c,i}^{t,m}, \av_{c,i}^{t,m}).
    \end{cases} 
\end{equation}
Let $\xv_{c,i}^{t,0} = \xv_{c,i}^{t}, \xv_{c,i}^{t,m} = \xv_{c,i}^{t+1}$, we can rewrite \eqref{eqn:step3} as
\begin{equation} \label{eqn:step3_2}
\begin{cases}
      \xv_{c,i}^{t+1} = \xv_{c,i}^{t} - \eta_t \sum_{\mv=0}^{h-1} \tilde{\nabla}_x F_{c,i}(\xv_{c,i}^{t,m}, \av_{c,i}^{t,m}) \\
      \av_{c,i}^{t+1} = \av_{c,i}^{t} - \eta_t \sum_{\mv=0}^{h-1} \tilde{\nabla}_a F_{c,i}(\xv_{c,i}^{t,m}, \av_{c,i}^{t,m}) 
    \end{cases} 
\end{equation}
where $\eta_t$ is the learning rate at round t and $\tilde{\nabla} F_{c,i}(\xv_{c,i}^{t,m}, \av_{c,i}^{t,m})$ is the derivative of the local loss for a specific mini-batch $\zv \in \tilde{D_i}$:
 \begin{equation} \label{eqn:step3_3}
 \tilde{\nabla} F_{c,i}(\xv_{c,i}^{t,m}, \av_{c,i}^{t,m}) = \frac{1}{|\tilde{D_i}|} \sum_{\zv \in \tilde{D_i}}  \nabla l(\xv_{c,i}^{t,m}, \av_{c,i}^{t,m} ; \zv).
\end{equation}
For the server-side model update, the server performs feedforward, calculates the loss, and updates the model sequentially using the smashed data $g_{\xv_{c,i}}(\zv)$ from $N$ clients. This is an important new step, as it allows the server to treat $N$ clients as $N$ mini-batches of training data in its own training. 
We also remark that the order of clients does not matter in \texttt{CSE-FSL}, which has an important practical advantage that the server can operate in a ``first come first serve'' mode by immediately processing the smashed data uploaded from any client without waiting for others, hence improving the overall latency. 
Note that the server updates the model in every $h$ batches of data and we define the model after each update with corresponding smashed data from client $i$ as $\xv_{s,i+1}^{t+1}$. The server performs model updates according to
\begin{equation} \label{eqn:step3_4}
\xv_{s,i+1}^{t+1} = \xv_{s,i}^{t+1} - \eta_t \tilde{\nabla} F_{s}(\xv_{s,i}^{t+1}, \xv_{c,i}^{t,h-1}).
\end{equation}
Let $\xv_{s,0}^{t+1} = \xv_{s}^{t}$ and $\xv_{s,N}^{t+1} = \xv_{s}^{t+1}$, we can rewrite \eqref{eqn:step3_4} as
$\xv_{s}^{t+1} = \xv_{s}^{t} - \eta_t \sum_{i=0}^{N-1} \tilde{\nabla} F_{s}(\xv_{s,i}^{t+1}, \xv_{c,i}^{t,h-1})$ 
where
$
\tilde{\nabla} F_{s}(\xv_{s,i}^{t+1}, \xv_{c,i}^{t,h-1}) =  \frac{1}{|\tilde{D_i}|} \sum_{\zv \in \tilde{D_i}}  \nabla l(\xv_{s,i}^{t+1} ; g_{\xv_{c,i}^{t,h-1}}(\zv))
$. 

\mypara{Step 4: global aggregation.} We use $C$ to denote the periodicity of global aggregation. In the theoretical analysis, $C$ is with the unit of batches, i.e., global aggregation happens every $C$ mini-batches of training. We first focus on the special case of $C=1$, which means that the global aggregation happens after \emph{every} mini-batch SGD step. Before each aggregation, client $i$ uploads the updated client-side model and the auxiliary model to the server. Then the server aggregates the client-side model and the auxiliary network according to
 \begin{equation} \label{eqn:step4}
\begin{cases}
      \xv_{c}^{t+1} = \frac{1}{N} \sum_{i=0}^{N-1} \xv_{c,i}^{t+1}   \\
      \av_{c}^{t+1} = \frac{1}{N} \sum_{i=0}^{N-1} \av_{c,i}^{t+1}.
    \end{cases} 
\end{equation}
After repeating the overall procedure for $T$ global rounds, the final aggregated model is the concatenation of aggregated client-side model and the final server-side model, which is utilized at the inference stage for different tasks.

\subsection{Convergence Analysis}
We analyze the convergence of \alg with non-convex loss functions and IID datasets. We focus on the setting of full device participation and per-batch aggregation (i.e., $|D| = N, C=1$). The more general case (non-IID, partial clients, and $C>1$) are left for the journal version of this work.
\begin{Assumption} \label{ass:smoo}
The client-side and server-side loss functions are L-smooth:  $\| \nabla F_c(\xv;\zv)-\nabla F_c(\yv;\zv) \|\leq L \norm{ \xv-\yv }, \| \nabla F_s(\xv;\zv)-\nabla F_s(\yv;\zv) \|\leq L \norm{ \xv-\yv } $ for any $\xv, \yv \in \mathbb{R}^d$ and any $\zv\in \mathcal{D}$. 
\end{Assumption}
\begin{Assumption} \label{ass:bound}
The expected squared norm of stochastic gradients is uniformly bounded. For the client-side loss, we have
$\Expect\| \nabla l(\xv_{c,i}^{t,m}, \av_{c,i}^{t,m};\zv) \|^2 \leq G_1^2 $, for any m, $i \in [n],  t \in [T]$ and any $\zv\in \mathcal{D}$.. Similarly, considering the server-side loss: $\Expect\| \nabla l(\xv_{s,i}^t ; g_{\xv_{c,i}^{t,m}}(\zv)) \|^2 \leq G_2^2 $,
for any m, $i \in [n],  t \in [T]$ and any $\zv\in \mathcal{D}$.
\end{Assumption}
\begin{Assumption} \label{ass:lr}
The learning rates satisfy $\sum_{t} \eta_t = \infty$ and $\sum_{t} \eta_t^2 < \infty$.
\end{Assumption}

Assumptions \ref{ass:smoo} and \ref{ass:bound} are standard in the literature \cite{boyd2004convex, li2019convergence}, and we set diminishing step sizes $\eta_t = \frac{\eta_0}{1+t}$, which satisfy the conditions in Assumption \ref{ass:lr} \cite{robbins1951stochastic}. 

Lastly, in each global round $t$, the input distribution of a specific server-side model is determined by $\xv_{c,i}^{t,m} $ and $D_i$. Let $z_{c,i}^t = g_{\xv_{c,i}^{t,h-1}}(\zv)$ be the output of the $i$-th client-side model at global round t, following the probability distribution of $P_{c,i}^t(z)$. Here $P_{c,i}^t(z)$ is time-varying, and we let $P_{c,i}^*(z)$ be the output distribution of the $i$-th client-side model with $x_c^*$ and $D_i$. We also define the distance between these two distributions as $d_{c,k}^t = \int \norm{P_{c ,i}^t(\zv) - P_{c,i}^*(\zv)} \,d\zv$. Based on this time-varying distribution, Assumption \ref{ass:dis} below is specific to our setting. A similar assumption has been adopted in \cite{belilovsky2020decoupled} but in centralized setup.

\begin{Assumption} \label{ass:dis}
We assume that $\sum_i d_{c,i}^t < \infty$.
\end{Assumption}



Due to space limitation, we present the following main theoretical results without giving the detailed proofs. 

\begin{Proposition}
\label{prop:conv_c}
Let $\Gamma_T = \sum_{t=0}^{T-1} \eta_t$. The following inequality holds for the client-side model in \alg
\begin{align} \label{eqn:conv1}
\frac{1}{\Gamma_T} \sum_{t=0}^{T-1} \eta_t \Expect \qth{ \norm{\nabla F_c(\xv_c^t)}^2} 
 \leq & \frac{4 (F_c(\xv_c^0) - F_c(\xv_c^*))}{(2M-1)\Gamma_T} \nonumber \\
&+ \frac{2M^2 G_1^2 L}{(2M-1)\Gamma_T} \sum_{t=0}^{T-1} \eta_t^2.
\end{align}
\end{Proposition}

As $T$ increases, the right-hand side of Eqn.~\eqref{eqn:conv1} converges to zero following Assumption \ref{ass:lr}. 

\begin{Proposition}
\label{prop:conv_s}
The server-side model of \alg converges as:
\begin{align} \label{eqn:conv2}
\frac{1}{\Gamma_T} \sum_{t=0}^{T-1} \eta_t 
& \Expect \qth{ \norm{\nabla F_s(\xv_s^t)}^2}
\leq \frac{4 (F_s(\xv_s^0) - F_c(\xv_s^*))}{(2N-1)\Gamma_T} \nonumber \\
&+ \frac{4 G_2^2}{2N-1} \frac{1}{\Gamma_T} \sum_{t=0}^{T-1} (\eta_t \sum_{i=1}^{N}d_{c,i}^t + \frac{LN^2}{2} \eta_t^2).
\end{align}
\end{Proposition}

From Proposition \ref{prop:conv_s}, we conclude that the expected gradient norm accumulates around 0 at rate $\inf_{t \leq T-1} \Expect \qth{\norm{\nabla F_s(\xv_s^t)}^2} <= \mathcal{O}(\frac{1}{\Gamma_T}\sum_{t=0}^{T-1} \eta_t \sum_{i=1}^{N}d_{c,i}^t )$. Overall, Propositions \ref{prop:conv_c} and \ref{prop:conv_s} state that both client-side and server-side training losses converge in \texttt{CSE-FSL}.


\section{Experimental Results}
\label{sec:sim}
\subsection{Setup}
\label{sec:settup}

We have carried out experiments to evaluate \alg on two popular datasets: 
CIFAR-10 \cite{krizhevsky2009learning} (60,000 images with 10 classes) and F-EMNIST \cite{caldas2018leaf} (81,7851 images with 62 classes). 
\color{black}
For CIFAR-10, we report experimental results for IID datasets and full clients participation. To demonstrate the effectiveness of our method, we also report the results for non-IID datasets and partial clients participation for the F-EMNIST dataset.

We consider the following strategies in the experiments. (1)~\textbf{FSL\_MC}: FSL with multiple model copies at the server. (2)~\textbf{FSL\_OC}: FSL with only a single server-side model. However, we have observed in all experiments that directly using one server-side model fails to converge for all tasks. Therefore, the popular gradient clipping method \cite{pascanu2013difficulty} is used to handle the gradient explosions. 
(3)~\textbf{FSL\_AN} (the same method in \cite{han2021accelerating}): add the auxiliary network to the client-side model and keep multiple server-side copies in the server. Moreover, each client $i$ uploads the smashed data to the server in every batch. (4)~\textbf{CSE\_FSL h}: our proposed \alg with parameter $h$. 
\color{black}
Each client trains $h$ batches of data at the same time and then uploads the smashed data to the server once. In the experiments, the global aggregation happens after the training process of one epoch ($C=1$). All of the reported results are obtained by averaging over five independent runs.

\subsection{Accuracy Comparison}
\label{sec:results}
\color{black}
We show our results with two different independent variables: epochs and communication rounds. One epoch means that all the training samples are used once for training. Moreover, when client $i$ sends the smashed data to the server, it completes one communication round.

\begin{figure}[htbp]
\centering
\includegraphics[width=0.48\linewidth]{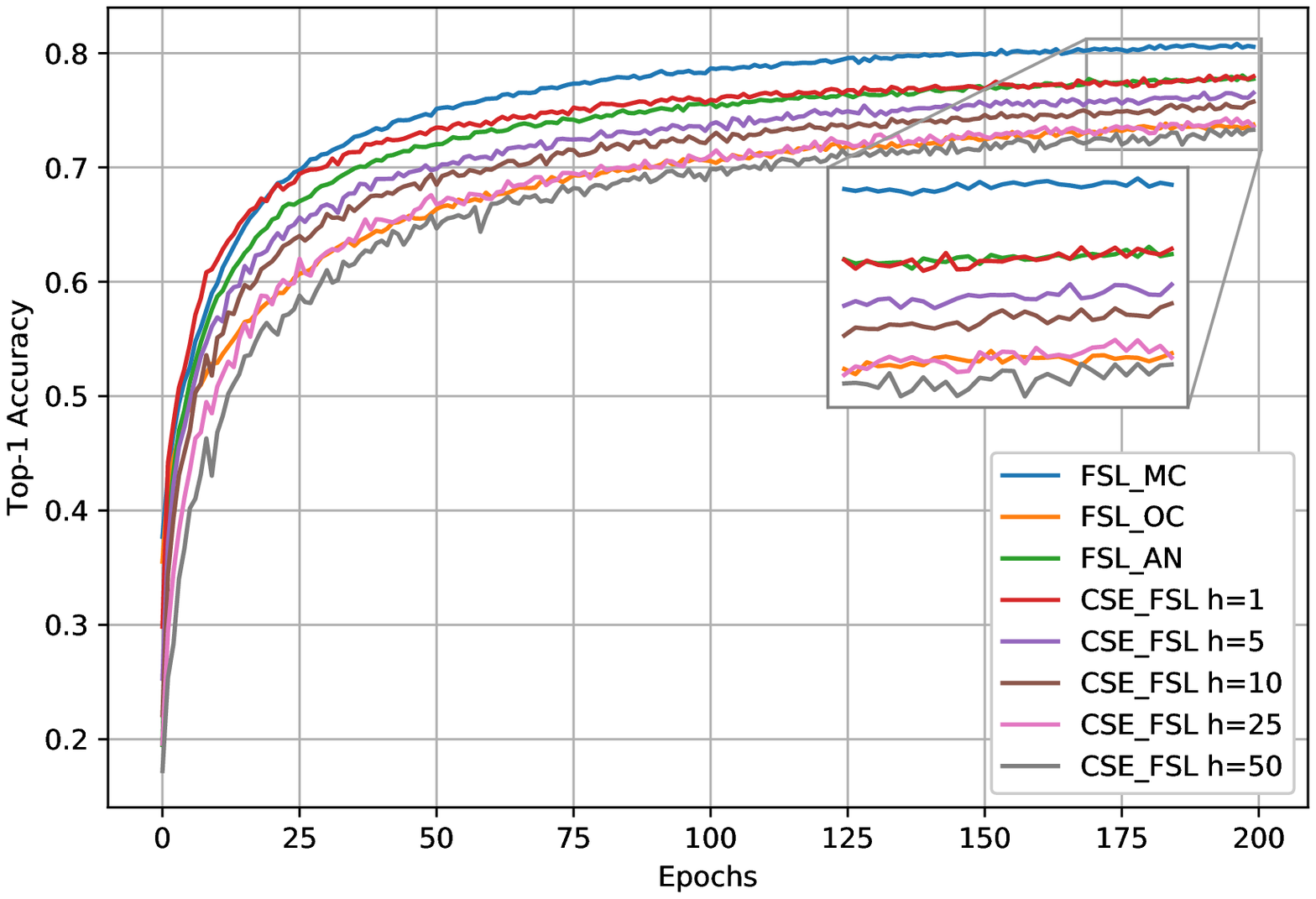}\hfill
\includegraphics[width=0.48\linewidth]{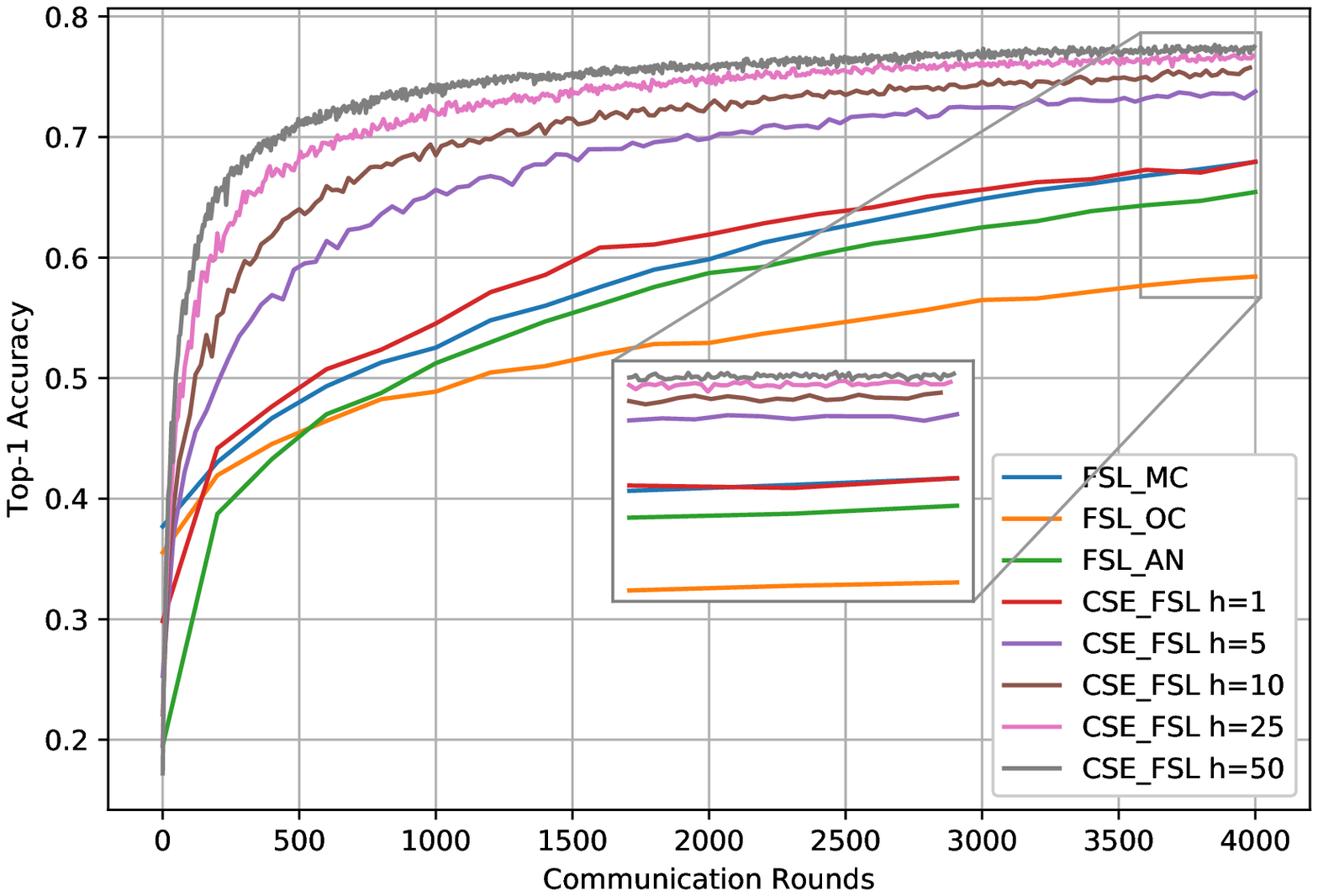}
\hspace*{25pt} \scriptsize (a) Accuracy vs. Epochs  \hspace{30pt} (b) Accuracy vs. Communication Rounds
\caption{\small CIFAR-10 Results with IID local datasets and full clients (5 clients) participation.}
\label{fig:cifar_5_clients}
\end{figure}

Fig.~\ref{fig:cifar_5_clients}(a) shows the performance of each method as a function of epochs in IID local datasets and full clients (5 clients) participation. 
In order to reduce the downstream communication load, FSL\_AN sacrifices a little performance compared to FSL\_MC. 
\color{black}
However, with only a very small size auxiliary network, 
CSE\_FSL with $h=1$ 
\color{black}
performs better than FSL\_OC even though the latter employs additional ML enhancements. This shows that the auxiliary network can solve the convergence problem when the server has only one server-side model. We also evaluate the top-1 test accuracy of our method with different parameters $h$, for example, $h=5$ means each client locally trains 5 batches of training data and then sends the smashed data to the server. It can be seen that our scheme with small $h$ performs better than large $h$ since more communications can help the server train more times and get high accuracy. Fig.~\ref{fig:cifar_5_clients}(b) compares the top-1 accuracy of each scheme during the same communication rounds.
It is easy to see that \alg has better performance than other methods in the same communication rounds, which indicates that our method still maintains the training performance without uploading smashed data in each mini-batch data samples. \color{black}
Overall, the top-1 test accuracy convergence accelerates with the increase of parameter $h$ due to more training steps in the same communication rounds.

\begin{figure}[htbp]
\centering
\includegraphics[width=0.48\linewidth]{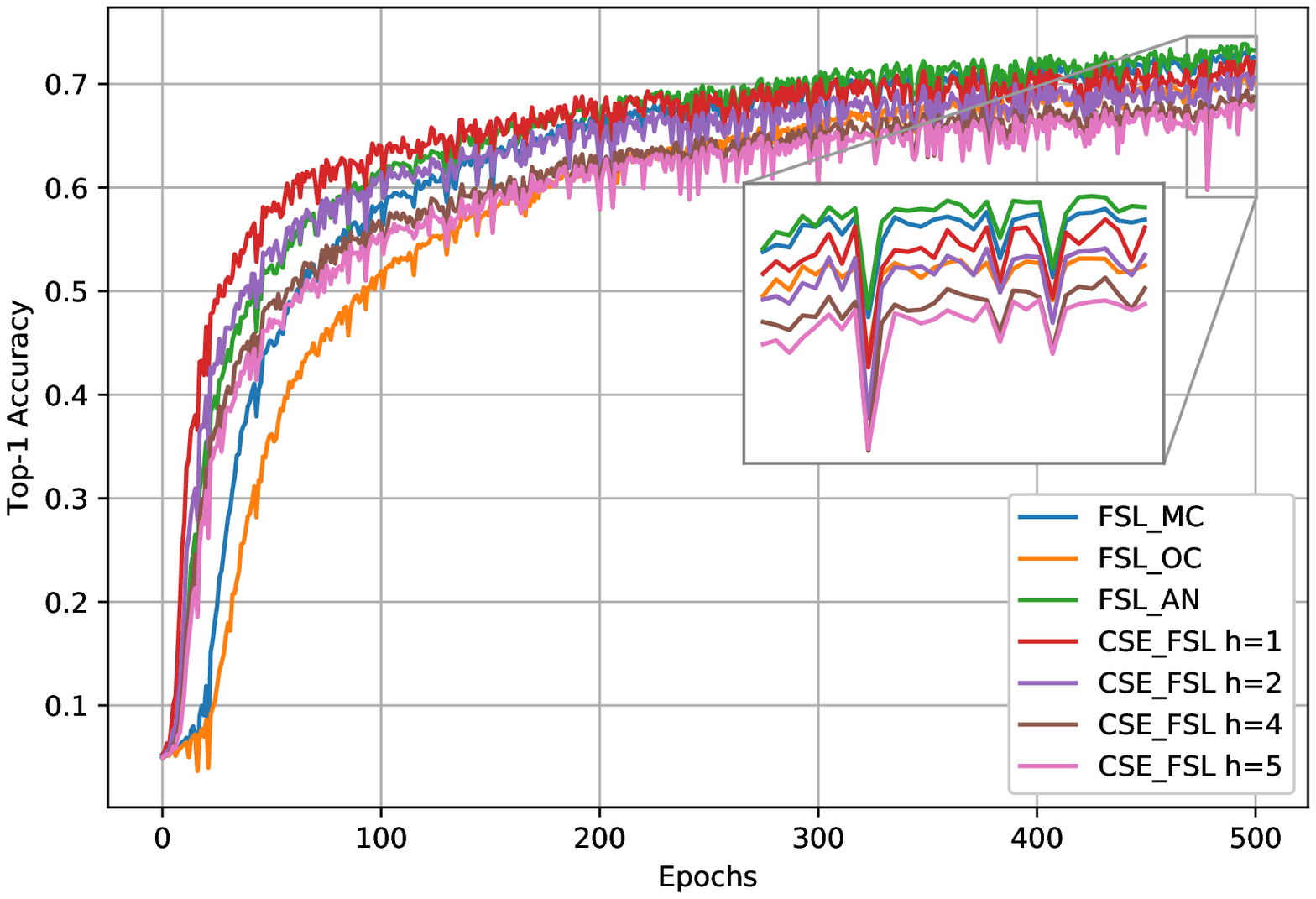}\hfill
\includegraphics[width=0.48\linewidth]{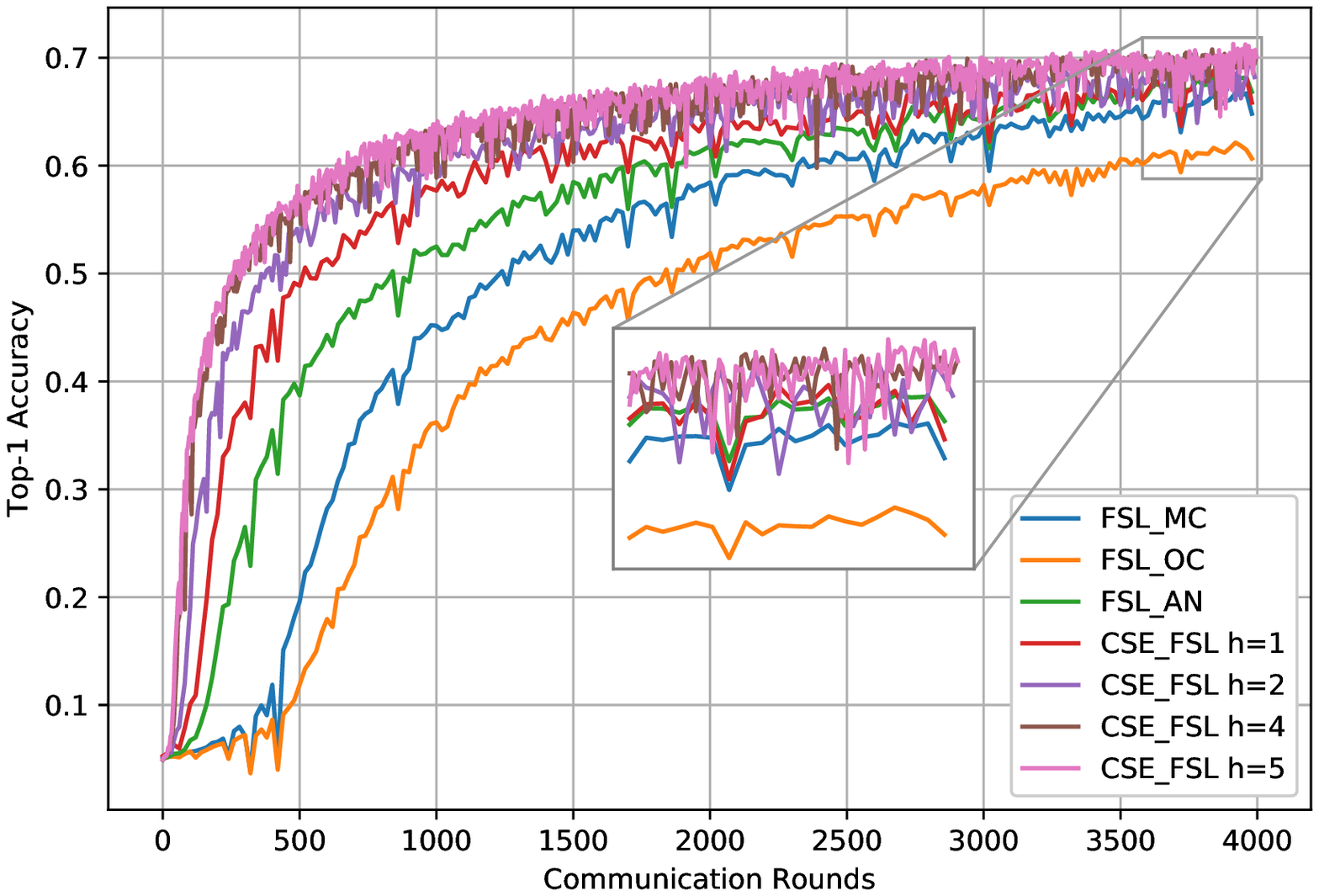}
\hspace*{25pt} \scriptsize (a) Accuracy vs. Epochs  \hspace{30pt} (b) Accuracy vs. Communication Rounds
\caption{\small F-EMNIST Results with non-IID local datasets and partial clients (5 clients) participation.}
\label{fig:femnist_5_clients_noniid}
\end{figure}

In the CIFAR-10 experiments, the entire training dataset is partitioned equally to the clients and all the clients join the ML training. In the second experiment for F-EMNIST dataset, we randomly choose partial clients in each global round. We similarly perform model training on the F-EMNIST dataset, and report the results in Fig.~\ref{fig:femnist_5_clients_noniid} for non-IID local datasets and partial clients participation. We see that FSL\_MC and FSL\_OC perform poorly despite exhaustive parameter tuning, while CSE\_FSLs again converge fast and achieve the 
top-1 accuracy of 65.85\%-70.09\%
\color{black}
after 4000 communication rounds.

\if{0}
\subsection{Effect of Auxiliary Network Structure}
\label{sec:aux}

In the CIFAR-10 experiments, the number of model parameters for the client-side model is 107,328, and the server-side model size is 960,970. Both CNN and MLP can be utilized for the auxiliary network, and the specific parameters are shown in Table \ref{tab:parameters}. Compared with the complete network size, the sizes of the auxiliary network are negligible. 

\begin{center}
\vspace{-0.1in}
\begin{table}[htb]
\caption{\small Parameters of Auxiliary network}

 \centering
 \setcellgapes{3pt}\makegapedcells
 \begin{tabular}{||l | c | c | c ||} 
 \hline \hline
 Method &\makecell{ Output channel\\number} &\makecell{Number of \\ parameters}  & \makecell{Percentage of \\ whole model}  \\ [0.5ex] 
 \hline\hline
MLP & N/A & 23,050 & 2.16\%\\
 \hline
CNN+MLP & 54 & 22,960 & 2.15\%\\
 \hline
CNN+MLP & 27 & 11,485 & 1.08\%\\
 \hline
CNN+MLP & 14 & 5,960 & 0.56\%\\
 \hline
CNN+MLP & 7 & 2,985 & 0.28\%\\
 \hline\hline
\end{tabular}
\label{tab:parameters}
\end{table}
\end{center}
\vspace{-0.2in}

\begin{figure}[htbp]
\centering
\includegraphics[width=0.48\linewidth]{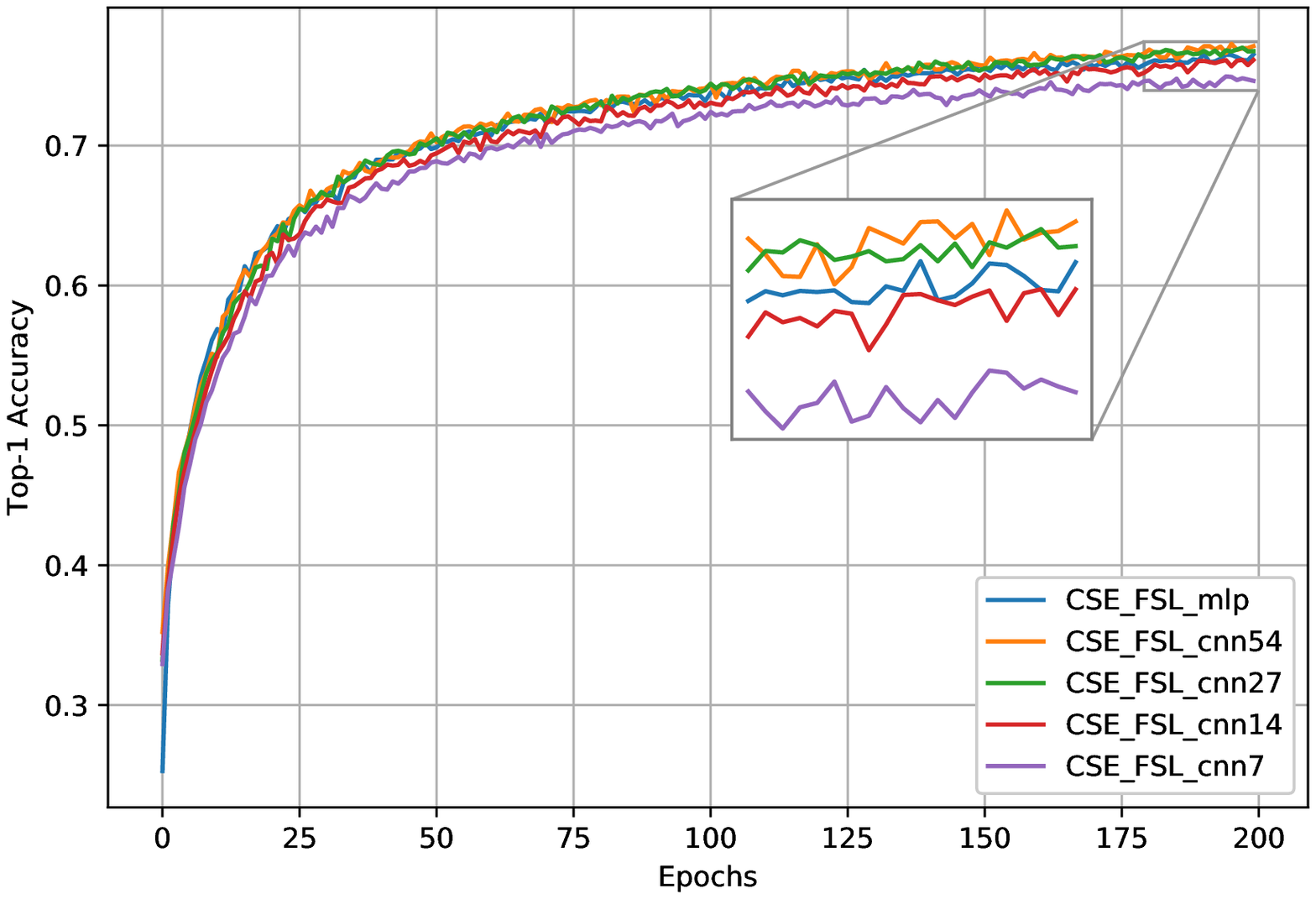}\hfill
\includegraphics[width=0.48\linewidth]{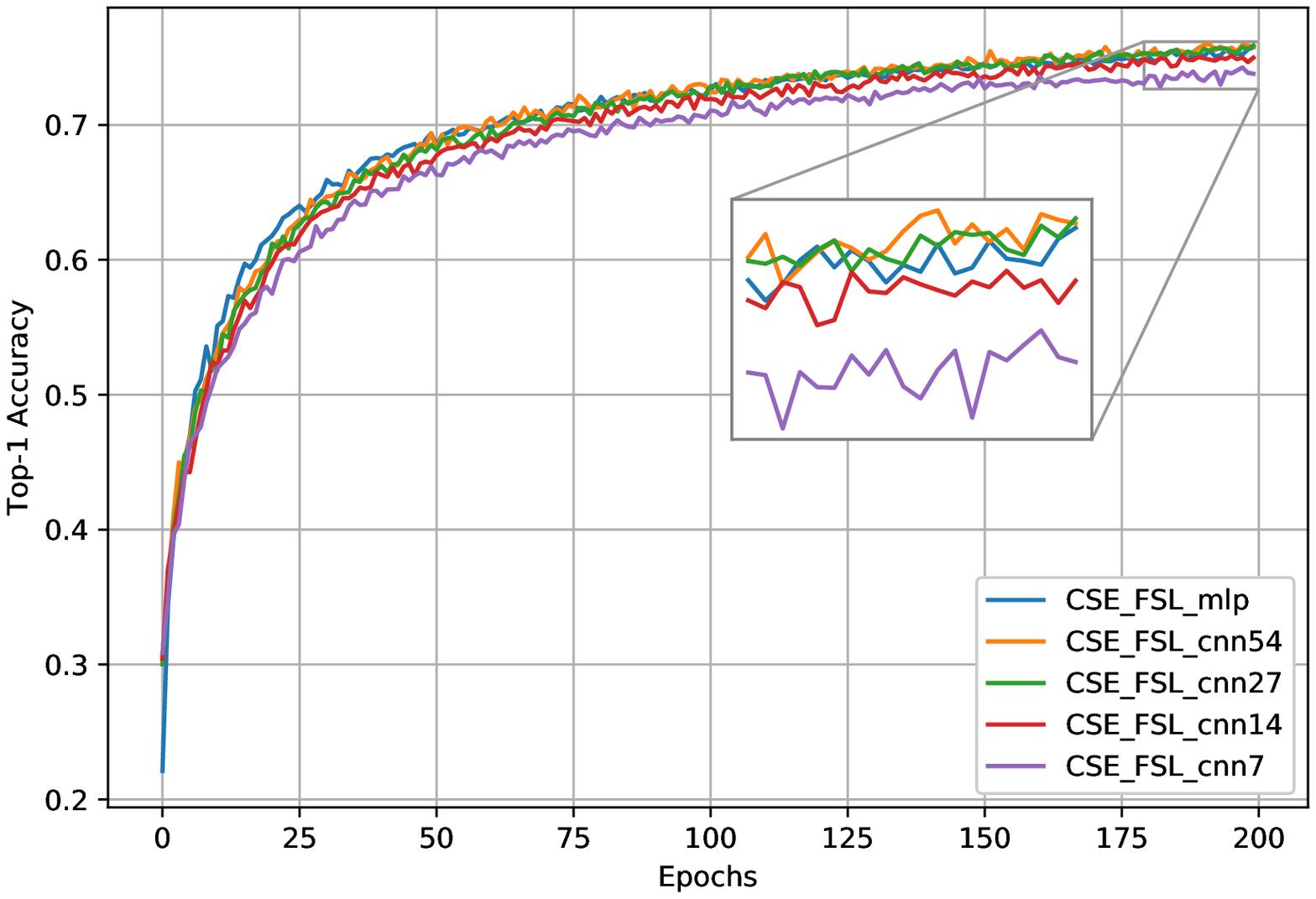}
\footnotesize (a) $h=5$ \hspace{100pt} (b) $h=10$
\caption{\small CIFAR-10 results with different auxiliary network structures.}
\label{fig:cifar_5_clients_aux}
\end{figure}

To further explore the impact of different network structures on the performance, we conduct experiments for IID local datasets and partial clients participation (5 clients). In general, we use $1 \times 1$ convolutional kernels to reduce the dimensionality of the filter space, which avoids the steep dimensionality drop like MLP. Primarily, we study the variation of top-1 accuracy with a decrease in the number of output channels of the CNN layer, and Fig.~\ref{fig:cifar_5_clients_aux} shows the performance when $h=5$ and $h=10$, respectively. As seen from both plots in Fig.~\ref{fig:cifar_5_clients_aux}, the CNN structure of the same size as the MLP can achieve the same level of accuracy and convergence speed and maintain unchanged even when the size is reduced by a factor of two. Accordingly, the auxiliary network with the CNN structure with 27 channels is much more storage-efficient than the initial MLP since it saves twice the storage, which is very important for low power devices.

\fi

\begin{center}
\vspace{-0.1in}
\begin{table*}[t]
\caption{\small Top-1 accuracy, communication load and storage comparison at 200 epochs on CIFAR-10 and 500 epochs on F-EMNIST.}

 \centering
 \setcellgapes{3pt}\makegapedcells
 \begin{tabular}{|l | c | c | c | c | c | c | c | c |} 
 \hline 
\multirow{2}{*}{\diagbox[height=23pt,width=7em]{Methods}{}}& \multicolumn{4}{c|}{CIFAR-10 IID} & \multicolumn{4}{c|}{F-EMNIST Non-IID} \\ \cline{2-9}

 & \multicolumn{2}{c|}{Accuracy (\%)} & Load (GB) & Storage ($M$)  & \multicolumn{2}{c|}{Accuracy (\%)} & Load (GB) & Storage ($M$) \\
\hline
FSL\_MC &  &\bf{80.55$\pm$0.21}  & 172.46 & 5.34  &   & 72.58$\pm$0.14 & 36.23 & 6.03\\
 \hline
FSL\_OC & & 73.74$\pm$0.23  & 172.46 & \bf{1.50} &  & 70.05$\pm$0.16  & 36.23 & \bf{1.28}\\
 \hline
FSL\_AN & & 77.75$\pm$0.10  & 93.96 & 5.46 & & \bf{73.26$\pm$0.18} & 51.06 & 8.89 \\
 \hline
\multirow{5}{*}{CSE\_FSL} 
& $h=1$ & 77.96$\pm$0.23  & \bf{86.80} & \multirow{5}{*}{1.61} & $h=1$ & 72.14$\pm$0.24  & \bf{28.16} & \multirow{6}{*}{4.14} \\
& $h=5$ & 76.52$\pm$0.41  & \bf{18.14} &   & $h=2$  & 70.63$\pm$0.24  & \bf{19.58} &   \\
& $h=10$ & 75.75$\pm$0.53  &\bf{9.55} &  &  $h=4$ & 68.77$\pm$0.25  & \bf{15.29} &\\
& $h=25$ & 73.57$\pm$0.60   & \bf{4.40} & & \multirow{2}{*}{$h=5$}  & \multirow{2}{*}{67.91$\pm$0.58}  &  \multirow{2}{*}{\bf{14.43}} & \\
& $h=50$ & 73.29$\pm$0.37  & \bf{2.69} &      & & & &\\
 \hline 
\end{tabular}
\label{tab:com_results}
\end{table*}
\end{center}
\vspace{-0.2in}

\subsection{Communication Load}
\label{sec:comm}

\begin{figure}[htbp]
\centering
\includegraphics[width=0.48\linewidth]{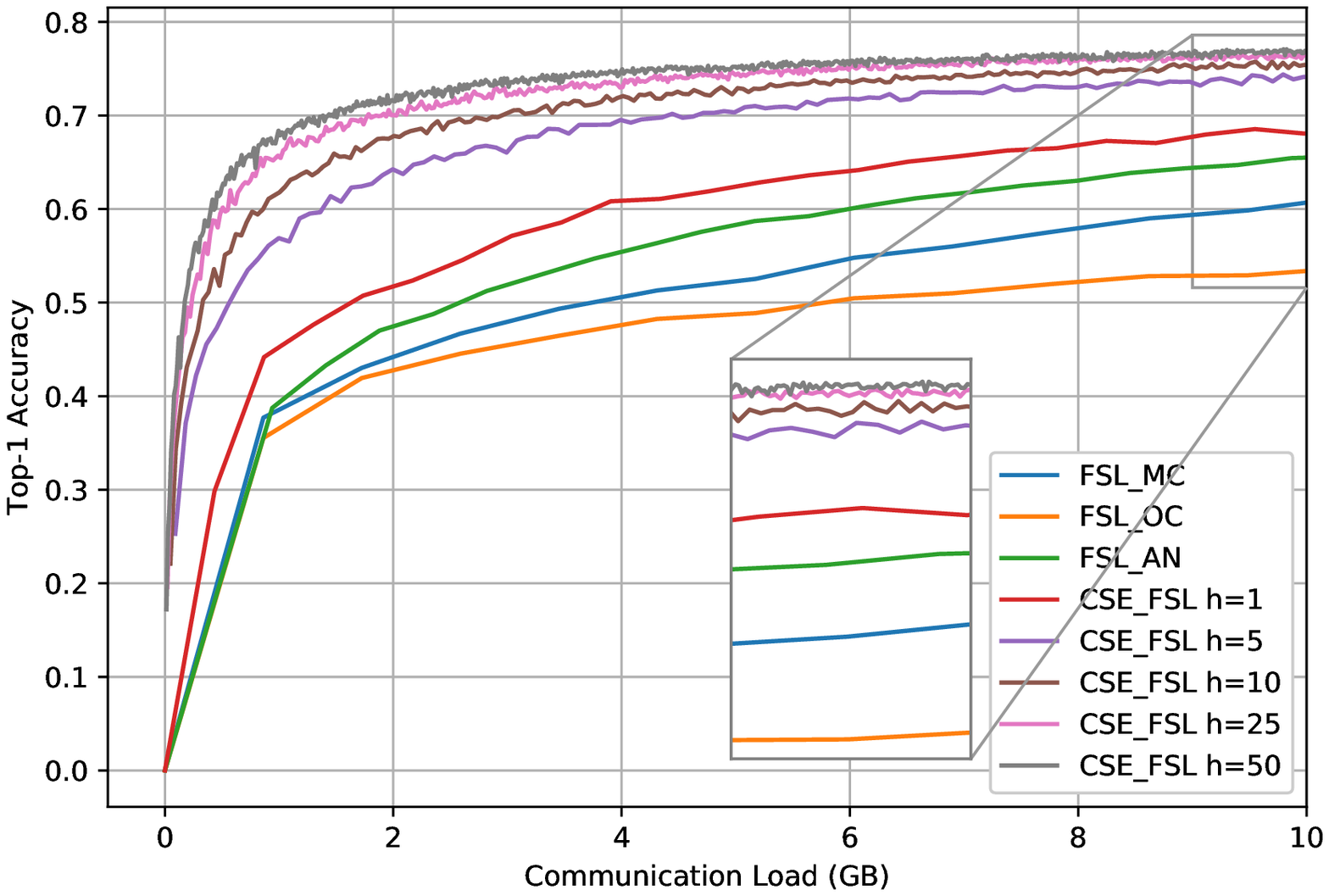}\hfill
\includegraphics[width=0.48\linewidth]{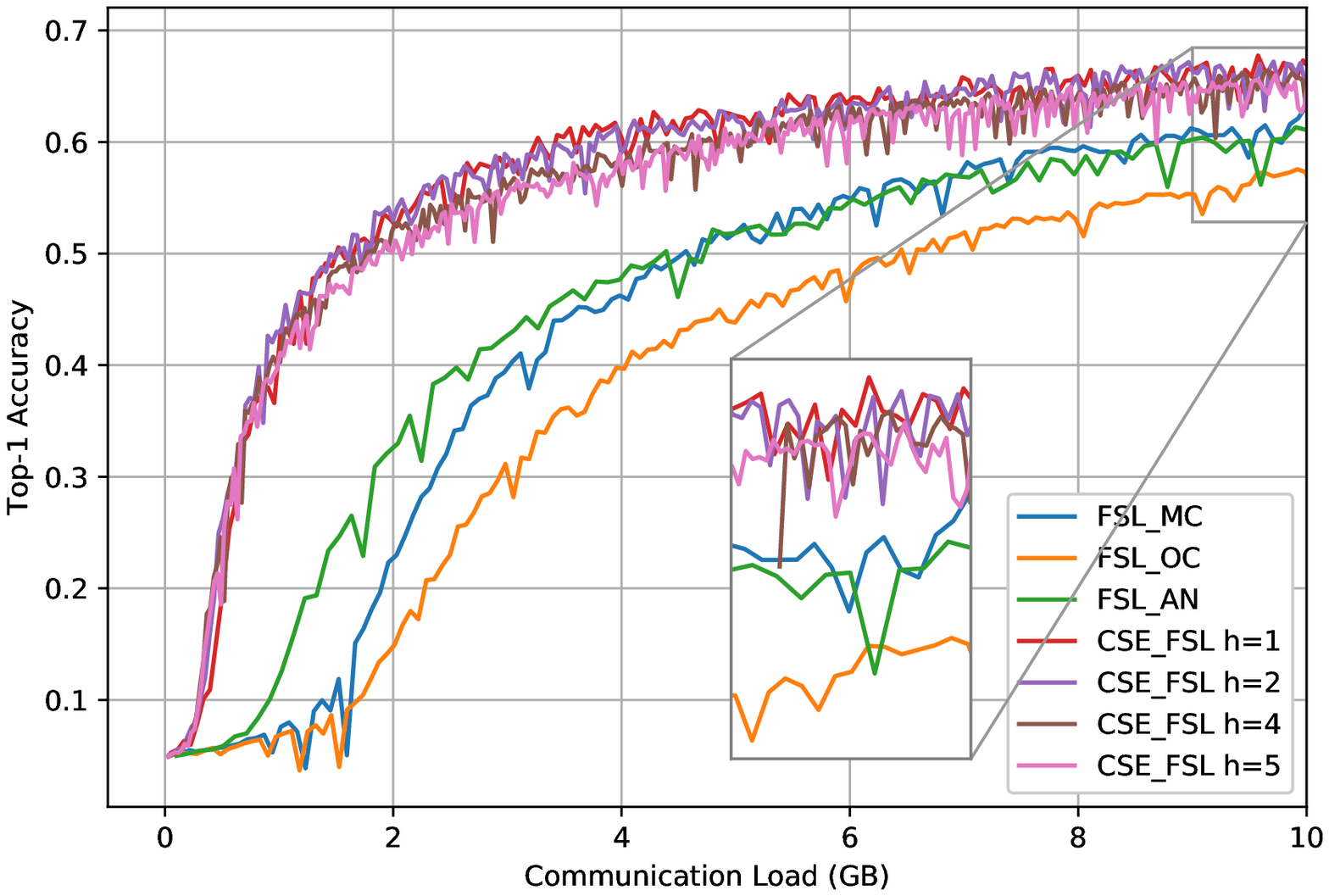}
\footnotesize (a) CIFAR-10 \hspace{70pt}  (b) F-EMNIST
\caption{\small Top-1 test accuracy versus communication load.}
\label{fig:comm}
\end{figure}

Fig.~\ref{fig:comm} shows the performance of each method as a function of communication load under CIFAR-10 with IID full clients participation and F-EMNIST with non-IID partial clients participation scenarios. From Fig.~\ref{fig:comm}(a), FSL\_AN performs better than FSL\_MC and FSL\_OC since no downlink communication for gradient transfer is required.
Compared with FSL\_AN, CSE\_FSL with $h=1$ achieves better top-1 accuracy while incurring the same communication load, which shows our proposed method \alg is much more communication-efficient.
\color{black}
Furthermore, CSE\_FSLs with large $h$ perform better and converge faster than with small $h$. This is because, in the CIFAR-10 task, the model is relatively simple, but each client has a large number of training samples, and consequently, the total reduction in smashed data uploads accounts for a larger proportion compared to client-side model transfer during the global aggregation.

From  Fig.~\ref{fig:comm}(b), we can see that all CSE\_FSLs with different $h$ can reach reasonable and good accuracy with less communication load. 
However, on the F-EMNIST dataset, CSE\_FSLs with larger $h$ do not outperform CSE\_FSL with $h=1$. 
\color{black}
This is due to the fact that the auxiliary network is too large (more complex model), and each client is assigned few training samples (partial client participation), so the reduction in smashed data is negligible. 
Therefore, CSE\_FSL with larger $h$ is more suitable for data-heavy clients or the size of the splitting layer is larger relative to the client-side model, because the communication load also depends on the number of mini-batches and the size of each batch transfer. 

\subsection{Storage and Comprehensive Analysis}
\label{sec:storage}

In the global aggregation steps, the server needs to aggregate the auxiliary networks (if applied) and client-side models. Therefore, the amount of storage is proportional to the number of clients. In addition to the above storage requirement, we also need to consider the server-side model. For example, the method of FSL\_MC has $N$ server-side models, while FSL\_OC and our method \alg only need to keep one server-side model during the whole training process. Here we consider to use the number of parameters to represent the model size, so the total storage is the sum of the auxiliary network, client-side model, and server-side model sizes. We report the storage comparison in Table \ref{tab:com_results}. FSL\_OC requires the minimal storage space because it has only one server model and no auxiliary network. On the other hand, FSL\_AN consumes huge storage space due to multiple server-side replicas and auxiliary networks. For our \alg, it saves more than 70\% storage space on CIFAR-10 and 53\% storage space on F-EMNIST than FSL\_AN.

In the same Table \ref{tab:com_results}, we summarize the top-1 accuracy, communication load, and storage space comparison under different methods on CIFAR IID and F-EMNIST Non-IID cases. It supplements the performance comparisons in Section \ref{sec:results} and Section \ref{sec:comm}. Compared with FSL\_MC, FSL\_OC reduces storage space, but generally degrades accuracy. Moreover, FSL\_AN improves communication efficiency but requires higher storage costs. Putting all aspects together, these results show that \alg consistently outperforms all other methods when considering the trade-off between top-1 accuracy, communication load, and storage space. In particular, \alg outperforms FSL\_AN with higher accuracy, lower communication load, and less storage cost on the CIFAR-10. 

\color{black}

\section{Conclusion}
\label{sec:conc}
We have proposed a novel federated split learning (FSL) scheme that is efficient in terms of both communication cost and storage space, and presented a theoretical analysis that guarantees its convergence. Novelty of \alg comes from using an auxiliary network to locally update client-side models, only keeping a single server-side model, and updating the server-side model sequentially leveraging the smashed data from all clients. 
With this method, we can reduce both upstream and downstream communication costs while saving storage space. 
\color{black}
Experimental results showed that \alg significantly outperforms existing FSL solutions with a single model or multiple copies in the server. \alg can be further enhanced by fine-tuning the structure of the auxiliary network and the amount of batch data to train a large-scale model in practical settings, especially in resource-limited devices.

\bibliographystyle{IEEEtran}
\bibliography{refs/reference}

\begin{thebibliography}{10}
\providecommand{\url}[1]{#1}
\csname url@samestyle\endcsname
\providecommand{\newblock}{\relax}
\providecommand{\bibinfo}[2]{#2}
\providecommand{\BIBentrySTDinterwordspacing}{\spaceskip=0pt\relax}
\providecommand{\BIBentryALTinterwordstretchfactor}{4}
\providecommand{\BIBentryALTinterwordspacing}{\spaceskip=\fontdimen2\font plus
\BIBentryALTinterwordstretchfactor\fontdimen3\font minus
  \fontdimen4\font\relax}
\providecommand{\BIBforeignlanguage}[2]{{%
\expandafter\ifx\csname l@#1\endcsname\relax
\typeout{** WARNING: IEEEtran.bst: No hyphenation pattern has been}%
\typeout{** loaded for the language `#1'. Using the pattern for}%
\typeout{** the default language instead.}%
\else
\language=\csname l@#1\endcsname
\fi
#2}}
\providecommand{\BIBdecl}{\relax}
\BIBdecl

\bibitem{mcmahan2017communication}
B.~McMahan, E.~Moore, D.~Ramage, S.~Hampson, and B.~A. y~Arcas,
  ``Communication-efficient learning of deep networks from decentralized
  data,'' in \emph{AISTATS}.\hskip 1em plus 0.5em minus 0.4em\relax PMLR, 2017,
  pp. 1273--1282.

\bibitem{gupta2018distributed}
O.~Gupta and R.~Raskar, ``Distributed learning of deep neural network over
  multiple agents,'' \emph{Journal of Network and Computer Applications}, vol.
  116, pp. 1--8, 2018.

\bibitem{thapa2020splitfed}
C.~Thapa, M.~A.~P. Chamikara, S.~Camtepe, and L.~Sun, ``Splitfed: When
  federated learning meets split learning,'' \emph{arXiv preprint
  arXiv:2004.12088}, 2020.

\bibitem{han2021accelerating}
D.-J. Han, H.~I. Bhatti, J.~Lee, and J.~Moon, ``Accelerating federated learning
  with split learning on locally generated losses,'' in \emph{ICML 2021
  Workshop on Federated Learning for User Privacy and Data
  Confidentiality}.\hskip 1em plus 0.5em minus 0.4em\relax ICML Board, 2021.

\bibitem{krizhevsky2009learning}
A.~Krizhevsky, ``Learning multiple layers of features from tiny images,''
  University of Toronto, Tech. Rep., April 2009.

\bibitem{caldas2018leaf}
S.~Caldas \emph{et~al.}, ``{LEAF}: A benchmark for federated settings,''
  \emph{arXiv preprint arXiv:1812.01097}, 2018.

\bibitem{boyd2004convex}
S.~Boyd and L.~Vandenberghe, \emph{Convex optimization}.\hskip 1em plus 0.5em
  minus 0.4em\relax Cambridge University Press, 2004.

\bibitem{li2019convergence}
X.~Li, K.~Huang, W.~Yang, S.~Wang, and Z.~Zhang, ``On the convergence of
  {FedAvg} on non-{IID} data,'' in \emph{International Conference on Learning
  Representations}, 2020.

\bibitem{robbins1951stochastic}
H.~Robbins and S.~Monro, ``A stochastic approximation method,'' \emph{The
  annals of mathematical statistics}, pp. 400--407, 1951.

\bibitem{belilovsky2020decoupled}
E.~Belilovsky, M.~Eickenberg, and E.~Oyallon, ``Decoupled greedy learning of
  {CNN}s,'' in \emph{International Conference on Machine Learning}.\hskip 1em
  plus 0.5em minus 0.4em\relax PMLR, 2020, pp. 736--745.

\bibitem{pascanu2013difficulty}
R.~Pascanu, T.~Mikolov, and Y.~Bengio, ``On the difficulty of training
  recurrent neural networks,'' in \emph{International conference on machine
  learning}.\hskip 1em plus 0.5em minus 0.4em\relax PMLR, 2013, pp. 1310--1318.

\end{thebibliography}

\end{document}